\documentclass[a4paper]{article}

\usepackage{multirow}

\usepackage{INTERSPEECH2022}
\usepackage{xcolor}

\title{Unsupervised Instance Discriminative Learning for Depression Detection from Speech Signals}
\name{Jinhan Wang$^1$, Vijay Ravi$^1$, Jonathan Flint$^2$, Abeer Alwan$^1$}
\address{
  $^1$Dept . of Electrical and Computer Engineering, University of California, Los Angeles, USA\\
  $^2$Dept. of Psychiatry and Biobehavioral Sciences, University of California, Los Angeles, USA}
\email{wang7875@g.ucla.edu, vijaysumaravi@ucla.edu, jflint@mednet.ucla.edu, alwan@ee.ucla.edu}

\begin{document}

\maketitle
\begin{abstract}

Major Depressive Disorder (MDD) is a severe illness that affects millions of people, and it is critical to diagnose this disorder as early as possible. Detecting depression from voice signals can be of great help to physicians and can be done without any invasive procedure. Since relevant labelled data are scarce, we propose a modified Instance Discriminative Learning (IDL) method, an unsupervised pre-training technique, to extract augment-invariant and instance-spread-out embeddings. In terms of learning augment-invariant embeddings, various data augmentation methods for speech are investigated, and time-masking yields the best performance. To learn instance-spread-out embeddings, we explore methods for sampling instances for a training batch (distinct speaker-based and random sampling). It is found that the distinct speaker-based sampling provides better performance than the random one, and we hypothesize that this result is because relevant speaker information is preserved in the embedding. Additionally, we propose a novel sampling strategy, Pseudo Instance-based Sampling (PIS), based on clustering algorithms, to enhance spread-out characteristics of the embeddings. Experiments are conducted with DepAudioNet on DAIC-WOZ (English) and CONVERGE (Mandarin) datasets, and statistically significant improvements, with $p$-value 0.0015 and 0.05, respectively, are observed using PIS in the detection of MDD relative to the baseline without pre-training. 


\end{abstract}

\section{Introduction}
Major Depressive Disorder (MDD) is one of the most severe chronic mental health disorders, characterised by persistent depressed mood, loss of interest in social activities, lack of energy, and even suicidal thoughts \cite{lopez2014major}. Around 4.4\% of the world's population  is estimated to suffer from depression \cite{world2017depression}. It is believed that the cause of MDD is multifactorial, which makes it unexplainable by conventional mechanisms and difficult to diagnose \cite{otte2016major}. Conventionally, diagnosing MDD is conducted by questionnaires  \cite{kroenke2009phq}. However, many subjective factors might cause mis-diagnosis \cite{khan2021automated}.  Therefore, automatic depression detection systems might help alleviate the above mentioned challenges.

There have been several studies focusing on automatic depression detection from signals, such as voice, video, electroencephalogram (EEG), etc. \cite{huang2019investigation, pampouchidou2015designing, wen2015automated, acharya2015computer}. Among those approaches, automatic depression detection based on speech signals draws more attention because speech signals can be an effective clinical marker for depression \cite{cummins2015review,lopez2014study}. Speech signals also prevail because they are relatively easy to obtain and analyze \cite{huang2019investigation}. Depressed individuals typically have slow, uniform, monotonous and/or hesitant voice characteristics \cite{liu2015detection, low2010influence}. There have been successful approaches to depression detection based on speech signals. Various features have been investigated, such as Mel Frequency Cepstral Coefficients (MFCCs) \cite{rejaibi2022mfcc,sharma2020detection}, Mel Filterbanks \cite{ma2016depaudionet}, vocal tract coordination (VTC) \cite{huang2020exploiting} and voice quality features \cite{afshan2018effectiveness,quatieri2012vocal, cummins2015review}. Aside from these handcrafted features, some approaches use raw audio signals \cite{bailey2020raw} in an end-to-end manner or high-level features extracted from pre-trained models, like x-vector~\cite{ravi2022fraug}, i-vector~\cite{afshan2018effectiveness}, NN2Vec \cite{salekin2018weakly}, and Depression audio embedding (DEPA) \cite{zhang2021depa}. In the back-end domain, researchers apply model architectures based on Convolutional Neural Networks (CNN) and Long-Short-Term Memory (LSTM) to capture spatial and temporal patterns of depression \cite{vazquez2020automatic, ma2016depaudionet}. Other additional mechanisms might be incorporated externally to improve model performance. For example, in \cite{vazquez2020automatic}, an ensemble method with a CNN has shown good robustness 
to acoustic variability, and in \cite{salekin2018weakly}  Salekin et al. reformulate the task into a weakly supervised learning problem. 

Unsupervised pre-training (UPT) is shown to be effective in various tasks given data scarcity, such as Automatic Speech Recognition (ASR) \cite{hsu2021hubert,fan2021bi}, and Speech Emotion Recognition (SER) \cite{li2021contrastive,jiang2020speech}. For SER tasks, Contrastive Predictive Coding (CPC) is applied by separating positive examples from negative examples though infoNCE minimization \cite{li2021contrastive}. Speech SimCLR is proposed to optimize contrastive loss between samples that are augmented \cite{jiang2020speech}. There have been a few studies that focused on UPT for depression detection. Most of these approaches use UPT models for feature extraction \cite{salekin2018weakly,zhang2021depa, sardari2022audio}. However, it's highly likely that the extracted features might lose important information for depression classification, since the pre-trained model is not optimized for depression detection. Hence, there is a domain discrepancy problem which might negatively affect performance. Inspired by a successful UPT method, Instance Discriminative Learning (IDL) \cite{ye2019unsupervised}, for image classification tasks, we propose a modified IDL approach for depression detection based on speech signals to extract augment-invariant and instance-spread-out embeddings in the pre-training stage. We investigate different sampling strategies for pre-training and find that the method that works the best preserves some speaker information. Additionally, a new sampling strategy, Pseudo Instance-based Sampling (PIS), is proposed to boost instance-spread-out characteristics. 



The remainder of this paper is organized as follows: Section 2 presents the IDL technique and the proposed PIS method. The experimental setup is described in Section 3,  followed by results and discussion in Section 4.  Section 5 concludes the paper.

\section{Instance Discriminative Learning}

\subsection{Pre-training Methodology}
\label{ssec:Pretraining}
\begin{figure}[t]
  \centering
  \includegraphics[width=8cm,height=4.5cm]{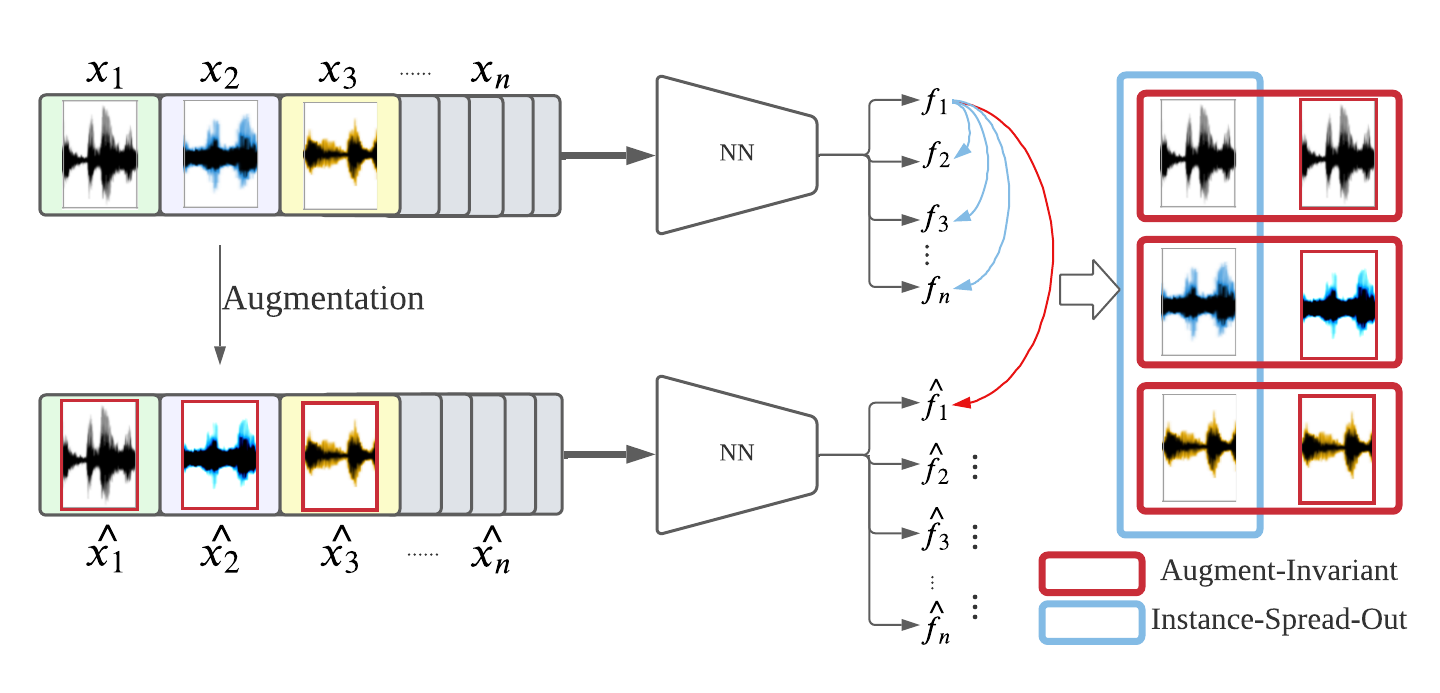}
  \caption{A schematic diagram of IDL pre-training. $x_i$ indicates an original instance and $\hat{x_i}$ represents its augmented version, $i = {1, 2, ..., n}$, where n is the batch size.  $f_i$ and $\hat{f_i}$ denote embeddings of $x_i$ and $\hat{x_i}$. Red and blue blocks are augment-invariant and instance-spread-out computations, respectively. }
  \label{fig:IDL}
\end{figure}

The schematic diagram of IDL \cite{ye2019unsupervised} is shown in Figure \ref{fig:IDL}. The assumption of IDL pre-training is that, the embedding of an instance and the embedding of its augmented object should be invariant, and embeddings of different instances should be spread-out. The embedding here is the output of the pre-trained model. In our case, all utterances are divided into segments with equal length, and each segment is taken as a distinct instance in pre-training. For each batch, \emph{n} segments are selected with pre-defined sampling strategies which are introduced in Section 2.3. For each segment $x_i$ in the batch, augmentation is applied and we obtain $\hat{x_i}$. Embeddings, denoted by $f_i$ and $\hat{f_i}$, are obtained by feeding the original segment and the augmented segment into the Neural Network (NN) module. To get an augment-invariant embedding, the probability that the augmented instance $\hat{x_i}$ being classified as instance $x_i$ is maximized, and is defined as:

\begin{equation}\label{eq:aug_inv}
    P(x_i|\hat{x_i}) = \frac{exp(<f_i,\hat{f_i} >/\tau)}{\sum_{k=1}^{n}exp(<f_k,\hat{f_i} >/\tau) }
\end{equation}
Embeddings should be instance-spread-out, where the probability that an instance $x_j$ being classified as another instance $x_i$, $j \neq i$, in a batch is minimized, and is defined as:
\begin{equation}\label{eq:ins_sp}
    P(x_i|{x_j}) = \frac{exp(<f_i,f_j >/\tau)}{\sum_{k=1}^{n}exp(<f_k,f_j >/\tau) }, j \neq i
\end{equation}
Here, probabilities are calculated as the ratios of exponential inner products of two embeddings scaled by a hyper-parameter $\tau$.  In all experiments , $\tau$ is empirically set to be 10. 

We assume that the probability of an instance $x_i$ being recognized as a different instance $x_j$ is independent for $j \neq i$. For each instance $x_i$, the joint probability that its augmented instance $\hat{x_i}$  can be recognized as $x_i$  and other instances $x_j$ cannot be recognized as $x_i$ is denoted as:
\begin{equation}\label{eq:Pi}
    P_i = P(x_i|\hat{x_i}){\prod_{j\neq i}(1 - P(x_i|x_j))}
\end{equation}
The NN module is then optimized by minimizing the average negative log-likelihood over a batch. Hence, the loss is :
\begin{equation}\label{eq:loss}
    L_{IDL} = -\frac{1}{n}(\sum_{i}log  P(x_i|\hat{x_i}) + {\sum_{i}\sum_{j \neq i}log(1 - P(x_i|x_j))})
\end{equation}

\subsection{Augment-Invariant Embeddings}
\label{ssec:Augmentation}
The first goal of IDL pre-training is to learn augment-invariant embeddings, corresponding to Equation \ref{eq:aug_inv}. Augmentation techniques are applied to instances in the batch during pre-training. However, for speech signals, effects of augmentation methods on depression status have not been fully explored. To mitigate potential negative effects of some augmentation methods that might change acoustic correlates of depression, such as pitch, formants, etc. \cite{cummins2015review}, augmentation methods used in this work are carefully chosen to be additive noise and volume perturbation at the signal level. For comparison, Vocal Tract Length Perturbation (VTLP) \cite{jaitly2013vocal}, as an augmentation technique that can change formant features, is also applied. All signal-level augmentations are conducted using open-source nlpaug package \cite{ma2019nlpaug}. In addition, to better analyze the effects of augmentation in the time and frequency domains separately, TM (time masking), FM (frequency masking) and SpecAugment (as a combination of TM and FM along with time-warping) \cite{park2019specaugment} are applied at the feature level, using mel-spectrograms.

\subsection{Instance-Spread-Out Embeddings}
\label{ssec:Sampling}
\subsubsection{Distinct speaker-based and Random Sampling}
Equation \ref{eq:ins_sp} attempts to achieve the goal of training embeddings to be instance-spread-out. Depending on different sampling strategies, the pre-training task has different implicit tendencies. In distinct speaker-based sampling (DS), it is guaranteed that each segment in a batch is from a distinct speaker. Hence, while the model is trying to spread out embeddings from different instances, it is also optimized to classifying different speakers. In other words, speaker information might be preserved in the embedding. The other sampling strategy is random sampling (RS), where segments in a batch are not constrained to be chosen from distinct speakers. If two samples in a batch are from the same speaker, they will still be classified into two instances during pre-training. Therefore, setting the RS strategy might eliminate speaker discriminative information from the embedding.


\subsubsection{Pseudo Instance-based Sampling}
Inspired by Hubert \cite{hsu2021hubert}, pseudo labels generated by clustering algorithms can reveal implicit correlation between hidden representations and underlying acoustic units in Automatic Speech Recognition (ASR) tasks. Embeddings trained from IDL might also have such a non-trivial correlation with depression status. The correlation can provide distinguishable characteristics across instances which may help with depression classification. Therefore, PIS is proposed to sample instances in a batch according to pseudo labels assigned by clustering algorithms instead of being speaker-dependent. Let $X$ denote all segments $X = [x_1, x_2, ..., x_n]$. At the first stage, the model is pre-trained using IDL with DS sampling strategy. Then, embeddings, $F = [f_1, f_2, ... f_n]$, are obtained by feeding $X$ into the best pre-trained model. Embeddings are clustered using a simple k-means model with $C$ cluster centroids. Corresponding pseudo labels, $\hat{Y} = [\hat{y_1}, \hat{y_1}, ..., \hat{y_n}]$, are assigned as cluster centroid variables, where $\hat{y_i} = {1, 2, ..., C}$. At the second stage, instead of sampling instances in a batch with DS, each instance is sampled from a distinct cluster to guarantee all samples in a batch have distinct pseudo labels. Therefore, $C$ is pre-determined to be the batch size in the second stage to guarantee all samples in a batch are from different clusters. 

\begin{table*}[t]
\centering
  \caption{Performance of IDL in terms of average F1-scores with different augmentation methods using distinct speaker-based sampling (DS) on DAIC-WOZ and CONVERGE test sets. Baseline is the experiments without pre-training. ND and D stands for non-depression and depression, respectively. In the Augmentation column, TM stands for Time-Masking, FM is Frequency-Masking, and SpecAug stands for SpecAugment. VTLP denotes Vocal Tract Length Perturbation. Noise and Volume stands for noise perturbation and volume perturbation, respectively.   The best F1-scores are boldfaced. $^*$ indicates that the change is not statistically significant.}
\label{tab:table1}
\begin{tabular}{ccccc|ccc}

  
   
Exp & Augmentation &  \multicolumn{3}{c}{DAIC-WOZ} & \multicolumn{3}{c}{CONVERGE} \\ \hline
& & F1 (avg)                   & F1(ND)                   & F1(D)     & F1 (avg)                   & F1(ND)                   & F1(D)               \\ \hline
\hline
Baseline& NA & 0.4258 & 0.5357 & 0.3158    & 0.7228 & 0.7101 & 0.7355    \\ \hline        
\multirow{3}{*}{IDL - Feature Level Aug}     & TM  &  \textbf{0.6458} & 0.8116 & 0.4800 & \textbf{0.7412}$^*$ & 0.7503 & 0.7321 \\ \cline{2-8} 
  & FM & 0.6103 & 0.7761 & 0.4444 & 0.7256 & 0.7343 & 0.7168\\ \cline{2-8} 
   & SpecAug  &  0.6189 & 0.8378 & 0.4000 & 0.7316 & 0.7432 & 0.7200\\ \hline 
\multirow{3}{*}{IDL - Signal Level Aug}   
   & VTLP & 0.5428 & 0.6984 & 0.3871 & 0.7197 & 0.7378 & 0.7015\\ \cline{2-8} 
& Noise &   0.6083 & 0.8000 & 0.4167 & 0.7293 & 0.7537 & 0.7048\\ \cline{2-8} 
  & Volume & \textbf{0.6458} & 0.8116 & 0.4800 & 0.7257 & 0.7453 & 0.7063\\ \hline


\end{tabular}
\end{table*}

\section{Experimental Setup}
For both pre-training and downstream datasets, we use 40 dimensional mel-spectrograms as input features, extracted with the Librosa library \cite{mcfee2015librosa} every 32ms with a 64ms Hanning window.
Experiments are conducted using PyTorch \cite{paszke2019pytorch}. 

\subsection{Datasets}
\label{ssec:Dataset}
\subsubsection{DAIC-WOZ}
The Distress Analysis Interview Corpus - Wizard of Oz (DAIC-WOZ) \cite{valstar2016avec} is an English dataset containing audio, text and video of interviews from 189 male and female participants, collected by a virtual interviewer. Audio files range between 7-33 min (16 min on average), with a non-depression vs depression ratio of 3:1. In this work, only recordings belonging to participants are extracted according to the groundtruth time stamps. Train, validation and test sets are partitioned in the same way as that provided in the dataset \cite{valstar2016avec}.


Librispeech \cite{panayotov2015librispeech} is used as the pre-training dataset for downstream tasks on DAIC-WOZ. It is the largest publicly available speech corpus in English with mostly reading style speech. Two subsets, training and validation, are partitioned with the ratio of 9:1 without speaker overlap, and the validation set is used to choose the best pre-trained model. 
\subsubsection{CONVERGE}
The Mandarin dataset we use is a part of the China, Oxford and Virginia Commonwealth University Experimental Research Genetic Epidemiology (CONVERGE) study~\cite{li2012patterns}. The dataset contains 391 hours of recordings from 7959 females with almost equal numbers of depression and non-depression classes. The sampling rate was 16kHz, and train, validation and test sets are split into 60\%, 20\% and 20\%. 


CN-Celeb \cite{fan2020cn} is chosen as the pre-training dataset for classification experiments conducted on CONVERGE. The dataset contains more than 130,000 utterances in Mandarin from 1000 Chinese celebrities across multiple genres. Training and validation subsets are partitioned in the same way as Librispeech. 

\subsection{DepAudioNet}
\label{ssec:DepAudioNet}
The back-end model we apply is DepAudioNet proposed in \cite{ma2016depaudionet, bailey2020raw} for depression detection from speech signals. 40-dimensional Mel-spectrogram features first pass through a one-dimensional convolution layer with a kernel size of 3 to capture short-term characteristics. Batch normalization is enabled after convolution. A max-pooling layer with a kernel size of 3 follows the convolution layer to capture mid-term features, with a dropout factor of 0.05 and an activation function as ReLU. Two LSTM layers and a fully connected layer activated by a Sigmoid with a hidden size of 128 are concatenated to generate binary predictions. 

For experiments on DAIC-WOZ, the model is pre-trained on Librispeech corpus. Utterances from each speaker are combined together and segmented into multiple fixed length segments with 120 frames each. The model is trained for 100 epochs with a batch size of 20. The learning rate is set to 1e-3, and the decay factor is set to 0.9 every two epochs. The model with the smallest validation loss is chosen to initialize the downstream model. For the depression classification task, to mitigate length variation and class imbalance, random cropping and random sub-sampling are applied \cite{ma2016depaudionet}. Each utterance is randomly cropped into a fragment with the same length as the shortest utterance, to mitigate any influence from longer objects. Then, each fragment is segmented with a fixed window length of 120 frames. A new subset is generated by sampling an equal number of depression and non-depression segments randomly without replacement. To fully utilize as many samples as possible, five individual models are trained using five randomly selected subsets for 100 epochs each, with a learning rate of 1e-3, and a decay factor of 0.9 every two epochs. Final predictions are obtained by averaging probabilities of the five models.

For CONVERGE experiments pre-trained on CN-Celeb, random cropping and sub-sampling are disabled because the dataset is balanced. One model is trained using all segments. The learning rate is empirically set to 1e-2. Other experimental configurations are the same as DAIC-WOZ experiments.

\section{Results and Discussion}
\subsection{Comparison of Different Augmentation Methods}
To investigate the effects of augment-invariant characteristics, different augmentation methods used in pre-training are compared while fixing the sampling strategy to be DS. F1-scores for the baseline system without pre-training, and IDL pre-training with various augmentation methods on DAIC-WOZ and CONVERGE test sets are shown in Table \ref{tab:table1}. All best F1-scores are statistically significant compared with the baseline unless specified.
In the feature-level augmentation experiments, the effect of time and frequency perturbation on depression status can be analyzed by perturbing the spectrogram in time or frequency at a time. 
Table \ref{tab:table1} shows that TM gives the best performance with a relative average F1-Score improvement of 51.67\% on DAIC-WOZ and 2.55\% on CONVERGE compared with the baseline system. FM is the worst among the three with a relative improvement of 43.33\% and 0.39\% on DAIC-WOZ and CONVERGE, respectively. Applying SpecAugment achieves an intermediate performance with a relative improvement of 45.35\% and 1.22\%, respectively. The results suggest that, augmenting the signal in the spectral domain, such as frequency masking, might result in loss of depression-specific information. The worst performance is observed for IDL with VTLP as the augmentation method, which also proves the hypothesis that modifications of spectral domain parameters can negatively affect depression classification. Because the spectrum is less affected by noise and volume perturbations compared with VTLP, moderate improvement can be observed on both datasets using these perturbations. Combination of augmentation methods are evaluated but performances are worse than using a single method.

\subsection{Sampling Strategy Comparison}
The instance-spread-out characteristics of the proposed pre-training method are explored by setting different sampling strategies with TM augmentation, since TM is the best augmentation technique found in the experiments reported in Section 4.1. Average F1-scores on DAIC-WOZ and CONVERGE test sets using different sampling strategies are shown in Table \ref{tab:table2}.

In Table \ref{tab:table2}, we observe that setting the batch sampling strategy to be DS yields relative improvements of 13.64\% and 4.79\% on DAIC-WOZ and CONVERGE, respectively, compared with using RS. Unlike RS, DS might preserve speaker information because embeddings of two speakers' instances are optimized to be spread-out, this observation implies speaker-discriminative information might be crucial in determining depression status. An ablation study is conducted in Section 4.3 to prove that speaker information is preserved using DS. 
\begin{table}[h]
\centering
  \caption{Average F1-scores of Baseline and unsupervised pre-training methods on DAIC-WOZ and CONVERGE. SS stands for sampling strategy. Best Results are boldfaced.}
  
\label{tab:table2}

\begin{tabular}{cccc}
 
F1(avg) & SS &  DAIC-WOZ & CONVERGE \\ \hline\hline
Baseline & - & 0.4258 & 0.7228 \\ \hline
CPC\cite{oord2018representation}  & -&0.5104   & 0.7371  \\  \hline
Speech SimCLR\cite{jiang2020speech} &- & 0.5747   & 0.7288   \\ \hline\hline
\multirow{3}{*}{IDL}&RS  & 0.5683  & 0.7073 \\ \cline{2-4}
&DS  & 0.6458   &    0.7412   \\ \cline{2-4} 

&PIS  & \textbf{0.6834} &  \textbf{0.7435} \\ \hline

\end{tabular}
\end{table}

In the IDL-PIS experiments, TM is also chosen for augmentation. PIS can further improve depression classification performance compared with DS, with relative improvements of 5.82\% on DAIC-WOZ and 0.31\% on CONVERGE. Improvements demonstrate that pseudo-labels generated by the clustering model provide a high correlation with depression status. 

As a comparison with other UPT methods, experiments with CPC \cite{oord2018representation} and Speech SimCLR \cite{jiang2020speech} without reconstruction loss are conducted and reported in Table \ref{tab:table2}. Results show that CPC and Speech SimCLR can perform better than the baseline system but not as well as the proposed IDL method. 

\subsection{Ablation Study on Speaker Classification}

We have shown that sampling instances in a batch using the DS sampling strategy during pre-training can help with the downstream depression classification task. To prove that speaker information is preserved in embeddings optimized by pre-training, speaker classification tasks with embeddings generated from downstream models as input are conducted on the DAIC-WOZ test set using a simple SVM classifier \cite{cortes1995support}. 30\% of the segments constitute the speaker classification test set. 

Table \ref{tab:table3} shows that, for IDL pre-training using TM with DS on Librispeech, without fine-tuning on the DAIC-WOZ, a 68.26\% speaker classification accuracy is achieved. The corresponding F1-score of 0.3472 is reasonable since the model hasn't been tuned for the depression task. The accuracy increases to 72.16\% after fine-tuning. This improvement can be explained by in-domain dataset adaptation through the downstream task. As a comparison, IDL (RS) only achieves 59.45\% speaker classification accuracy. The relative speaker classification accuracy degradation of 17.61\% from DS to RS proves that speaker information are better preserved using DS. Additionally, using PIS achieves a comparable speaker classification accuracy and an improved F1-score on depression classification compared with DS. This observation reveals that depression-specific characteristics can be preserved in embeddings trained using PIS, along with some speaker information. 
\begin{table}[t]
\centering
  \caption{Average F1-scores and Speaker Classification Accuracy (Spk Cls Acc) of Baseline and unsupervised pre-training on DAIC-WOZ. w/o ft stands for no fine-tuning}
  
\label{tab:table3}

\begin{tabular}{lccc}
 
 Exp       &  F1(avg) & Spk Cls Acc(\%) \\ \hline\hline
 Baseline & 0.4258 & 21.98 \\ \hline
 IDL (DS w/o ft)  & 0.3472   & 68.26  \\  \hline
 IDL (DS ) & 0.6458   & 72.16   \\ \hline
 IDL (RS) & 0.5683   &    59.45   \\ \hline   
 IDL (PIS)  & 0.6834 &  71.60 \\ \hline  
 
\end{tabular}
\end{table}

\section{Conclusion and Future Work}
In this paper, a modified UPT approach, IDL, is proposed to learn augment-invariant and instance-spread-out embeddings for depression detection tasks on DAIC-WOZ and CONVERGE datasets. Different augmentation techniques are compared in terms of augment-invariant characteristics. Results show that TM yields the best performance among all augmentation methods. For learning instance-spread-out embeddings, different sampling strategies, DS and RS are investigated and compared. Results show that preserving speaker information in the embedding using DS might help with depression classification. We also propose a new sampling strategy, PIS, to generate pseudo labels based on clustering, to reveal a deeper correlation between embeddings and depression status. Compared with the baseline, DS and RS, the proposed approach achieves significant improvements. In future work, we will investigate context-based, in addition to speaker-based, sampling strategies and apply IDL and PIS to other low-resource tasks.

\section{Acknowledgements}

This work was funded in part by the National Institutes of Health under the award number R01MH122569-Combining Voice and Genetic Information to Detect Heterogeneity in Major Depressive Disorder.

\bibliographystyle{IEEEtran}

\bibliography{mybib}


\end{document}